# A Phase Shift and Sum Method for UWB Radar Imaging in Dispersive Media

Wenyi Shao, *Senior Member, IEEE*, Todd R. McCollough, *Member, IEEE*, and William J. McCollough, *Member, IEEE*

*Abstract*—A phase shift and sum (PSAS) algorithm to image objects in dispersive media is presented. The algorithm compensates the phase shift of the scattered field from the receiver to the source for each frequency component in an ultrawideband (UWB) and then integrates all the frequency responses. This method resolves the multispeed and multipath issue when UWB signals propagate in dispersive media. In addition, a multipath effect due to refraction on a curved boundary is also explored. By collecting data using a customized microwave measurement system of two different objects placed in a plastic graduated cylinder filled with glycerin, along the measured dielectric parameters of glycerin (a dispersive medium), high-quality reconstructed images are formed using PSAS. Quantitative and qualitative comparisons with two other traditional time-shift radar-based microwave imaging algorithms for the same objects under test demonstrate the advantages of PSAS.

## I. Introduction

ULTRA-WIDEBAND (UWB) radar imaging aims to identify the presence and location of scatters such as underground objects (by ground-penetrating radar) [1], objects behind walls (by through-wall radar) [2], [3], breast tumors [4], [5], and brain stroke regions [6], [7]. A variety of image formation algorithms have been proposed over the last decade. A classic method is the confocal microwave imaging (CMI) algorithm which adopts a delay-and-sum (DAS) beamforming technique [8], [9]. Compared with DAS, improved performance of clutter rejection has been achieved by delay-multiply-and-sum (DMAS) [10], modified-weighted DAS (MWDAS) [11], microwave power imaging (MPI) [12], filtered DAS (FDAS) [13], [14], and robust artifact resistant (RAR) [15]. These methods rely on correctly evaluating the time shift (TS) of a UWB signal between an antenna and a focal point. However, in many applications [1]–[7], objects under test are covered by or within dispersive media leading to inaccurate results.

There are three primary reasons methods [8]–[15] relying on TS can lead to inaccurate results when dealing with the dispersive effect. First, different frequency components of a UWB signal may propagate with different velocities and travel in different paths from one medium to another. In such scenarios, TSs are unable to be accurately estimated. Second, in near-field problems, the dimension of the antenna (especially with an advanced, high-gain antenna) may be comparable to the electrical distance between the object and the antenna, leading to difficulty in determining the correct distance value to use in the TS calculation. Although the antenna's phase center represents a point position which can ease the distance calculation, the phase center of a UWB antenna is also frequency dependent. Finally, dispersion distorts the UWB pulse. A bandwidth reduction can be observed when a UWB signal propagates in a dispersive lossy medium for a certain range, because high-frequency components attenuate faster than lower frequency components, which ultimately degrades imaging resolution. Thus, the TS evaluation these prior methods (i.e., CMI, DAS, DMAS, MWDAS, FDAS, MPI, and RAR) rely on do not allow for accurate estimations of the presence and location of objects.

Instead of evaluating the TS, the phase confocal method (PCM) [16] evaluates a phase shift (PS) in the frequency domain between an object and an antenna. Shao *et al.* [16] validated the PCM method by localizing objects exposed in air, discarding amplitude information, and using only phase information of a scattered field to form images. The algorithm presented in this paper inherits the concept of replacing TS by PS from PCM but will use both phase information and amplitude information to detect objects within dispersive media. In the PS and sum (PSAS) method, each frequency component in the UWB scattered signal is individually processed for PS compensation and amplitude-decay compensation. Then, the PS frequency responses are integrated over the spectrum, and the result is converted to a pixel value at the focal point to form an image. Since the present method applies a frequency-wise mode, the previously described challenges inherent in time-domain UWB approaches are overcome. More specifically, the advantages of PSAS are as follows.

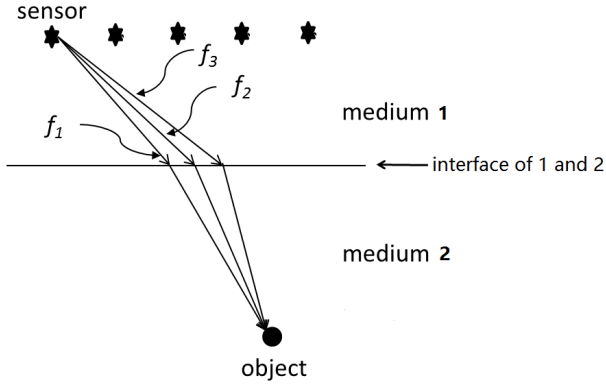

Fig. 1. Geometry of the microwave detection problem. Medium 1 and medium 2 have different dielectric properties. Medium 2 is typically lossy and dispersive and contains the object to be detected. A UWB signal having many frequency components $f_1$, $f_2$, $f_3$, …takes multiple paths to travel between the sensor and the object.

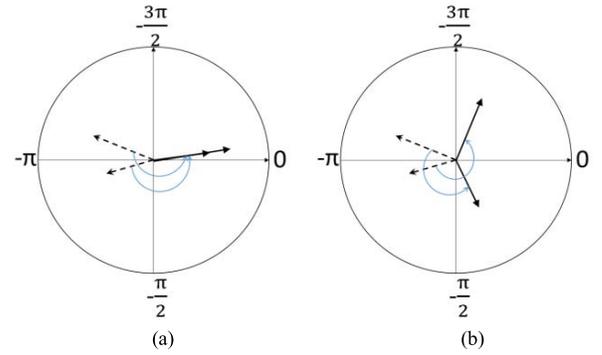

Fig. 2. (a) Two signals phase shifted to the targets' position produce a large-magnitude vector when summed. (b) Two signals phase shifted to a nontarget position produce a small-magnitude vector when summed.

1) PSs between the antenna and focal points can be accurately evaluated because each frequency component in the UWB spectrum is individually treated.
2) The frequency-dependent phase centers of the UWB antenna can be implemented to ease the PS estimation.
3) Amplitude-decay in the lossy media is compensated for each frequency component, which relieves the UWB waveform distortion.

Therefore, the PSAS algorithm promises to resolve the dispersion effect and provide better quality images for objects immersed in dispersive media.

The details of the PSAS algorithm are described in Section II. Meanwhile, a nonfrequency-selective multipath effect due to refraction on a curved interface is also discussed in Section II. An experimental setup to validate the PSAS algorithm is presented in Section III. A comparison of imaging results between PSAS and two other algorithms relying on TS is presented in Section VI, which specifically is evaluated in terms of three image quality parameters: signal-to-noise ratio (SNR), image contrast, and relative difference between a reconstructed image and an ideal object profile. The PSAS algorithm and the two other algorithms are also evaluated in terms of the ability of missing weak scatters when multiple scatters exist and the extent of distorting the object shape. Finally, conclusions are offered in Section V.

## II. FORMULATION OF PSAS ALGORITHM

The geometry of the microwave detection problem is simplified and shown in Fig. 1. Sensors (antennas) are placed in medium 1 (which could be air or a coupling liquid for biomedical examinations) to transmit a pulse signal and receive the scattered field of an object. Medium 2 is lossy and dispersive and contains the object to be detected (where the object has different dielectric properties than that of medium 2). Due to the dispersive property of medium 2, different frequency components in the UWB pulse propagate with different velocities in medium 2, take different paths to reach the object, and their amplitude decay also varies. Conventional time-domain radar-based methods [8]–[15] usually adopt the speed and path at a certain frequency in UWB to evaluate the time delay between the antenna and target, which is not accurate. PSAS processes all frequency components in the UWB signal individually. A PS between the antenna and a focal point is processed at each frequency, and finally contributions of all frequency components are summed to reveal the effect of the entire UWB signal. Assuming $M$ transmitters (TXs) take turns transmitting the UWB signal, and in each transmission, echoes are received by $N$ receivers, measurements can be carried out by a frequency sweep or performed in the time domain and then converted to the frequency domain by a Fourier transform. As the first step of PSAS, PS based on the propagation path from TX to focal point in medium 2 and then to the receiver is estimated by

$$\emptyset = 2\pi \left( \frac{d_{1T}}{\lambda_1} + \frac{d_{2T}}{\lambda_2} + \frac{d_{3R}}{\lambda_2} + \frac{d_{4R}}{\lambda_1} \right) \quad (1)$$

where $d_{1T}$, $d_{2T}$, $d_{3R}$, and $d_{4R}$ represent the distance the incident wave travels from TX to interface in medium 1, from the interface to the focal point in medium 2, the scattered wave travels from the focal point to the interface in medium 2, and from the interface to the receiver in medium 1, respectively. The phase compensation process is carried out as if the vectors shown in Fig. 2 turned counterclockwise. Consider two harmonic signals (vectors) acquired by the antenna located in two different positions. In Fig. 2(a), PS signals to a position most likely presenting a target will have consistent phase. Hence, the sum of the PS signals will produce a large-magnitude vector. In contrast, Fig. 2(b) represents two signals shifting back to a position where most likely no target presents. Thus, the sum of two signals will produce a small-magnitude vector because of out of phase. The magnitude of the sum will contribute to calculating the pixel value of an image. In practice, there are $M \times N$ signals to shift and sum at each frequency.

The magnitude of the sum described above represents the power density at a certain frequency. An integration of the power density over the effective bandwidth $f_H - f_L$ ($f_L$ and $f_H$ are the low-end and high-end frequency of a UWB signal, respectively), as shown in Fig. 3, represents the power focused on the focal point. Mathematically, this process is

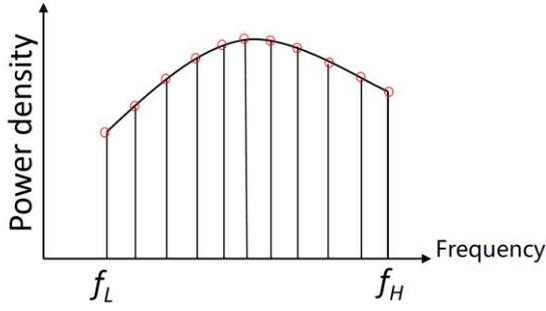

Fig. 3. Sum of $M \times N$ vector signals at a frequency is taken as the power density for the frequency; then, an integral over the bandwidth $f_H - f_L$ represents the power of the UWB signal.

written by

$$P = \int_{f_L}^{f_H} \left| \sum_{n=1}^{M \times N} \frac{|\vec{R}| e^{j\tilde{k}\vec{R}} \cdot V_n(f, \emptyset)}{\sqrt{G_t(x, y, f) \cdot G_r(x, y, f)}} \right|^2 df \quad (2)$$

where $V_n(f, \emptyset)$ is a measured scattered field containing phase information at frequency $f$ ($f_L < f < f_H$); $G_t(x, y, f)$ and $G_r(x, y, f)$ represent the gain of TX and receiver antenna representatively at the focal point $(x, y)$ for frequency $f$, $G_t(x, y) = G_r(x, y) = 1$ when the antennas involved are isotropic; $\tilde{k}$ is the complex wavenumber; and the function of $e^{j\tilde{k}\vec{R}}$ is more than a phase compensator like in a dispersive synthetic aperture radar (SAR) system. In our method, $e^{j\tilde{k}\vec{R}}$ compensates both PS and amplitude decay due to absorptions in the lossy medium. Assuming the signal travels in two media, e.g., medium 1 and medium 2 as shown in Fig. 1, then

$$e^{j\tilde{k}\vec{R}} = e^{j[\tilde{k}_1(d_{1T} + d_{4R}) + \tilde{k}_2(d_{2T} + d_{3R})]} \quad (3)$$

where $\tilde{k}_1$ and $\tilde{k}_2$ are wave numbers for medium 1 and medium 2, respectively. If medium 1 is lossless, then $\tilde{k}_1$ becomes a real number. Medium 2 is usually lossy, i.e., $\tilde{k}_2 = k_2 - j\kappa$. Thus, the amplitude attenuation is compensated by $e^{\kappa(d_{2T}+d_{3R})}$. $|\vec{R}|$ and $G(x, y, f)$ affect the amplitude compensation as well, which might be less important in SAR imaging but is not trivial on some occasions, which will be elaborated later in this section. An integration over $f_H - f_L$ would then give the power of the UWB signal that will be linearly converted to a pixel value in the image.

Note that variable $\vec{R}$ in (2) and (3) is also a function of frequency. In conventional time-domain radar-based methods, $\vec{R}$ was considered as the same for all frequency components, meaning that it does not consider the variation of the phase center of a UWB antenna, the variation of the index of refraction, and the variation of loss (in a dispersive lossy medium) for different frequency components. Equation (3) provides a frequency-selective compensation which effectively solves the multispeed and multipath effect due to dispersion in medium 2. In addition, there is a second kind of multipath effect which is not due to the dispersion, but the refraction on a curved interface between medium 1 and medium 2. We can use harmonic wave to investigate the second kind of multipath. Hereafter, the first kind of multipath is named the "dispersive multipath," and the second kind "monochromatic multipath."

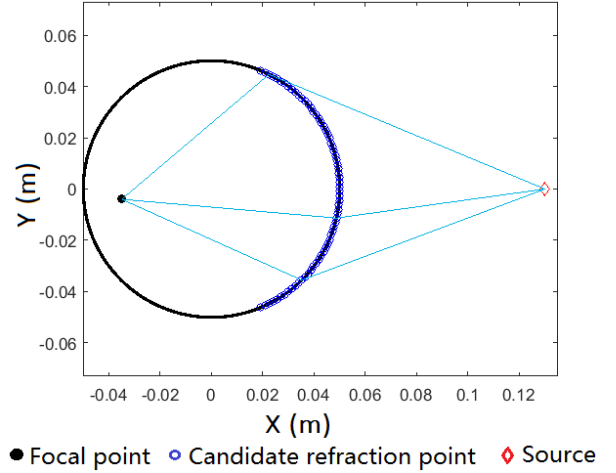

Fig. 4. Cross section of a 10-cm-diameter cylinder illuminated by a source placed 13 cm to the right of the center. There are three paths, in which the wave propagates from the source to the focal point inside the cylinder.

When refractions occur in the path, the least-time method is often used alone to find refraction on the interface of two media [6], [17], [18], by which the propagation path is revealed. However, the least-time method is a misunderstanding of Fermat's principle. A more modern interpretation of the principle is that a wave takes a path of stationary time length with respect to the variations of the path. Here, "stationary" means that the first derivative is equal to zero.

Consider a 2-D geometry as shown in Fig. 4. A cross section of a cylinder having a circular boundary (might be the simplest curved boundary in practice) with diameter of 10 cm is the region of interest (ROI), which is assumed lossless for easy analysis, with assigned permittivity $\varepsilon_r = 6.5$ at a frequency of 4.5 GHz. Outside of the cylinder is air. A TX source is placed 13 cm to the right of the cylinder's center. To find out the path when the wave propagates from TX to focal point ($x = -0.035$ m, $y = -0.003$ m) shown as a black point in the cylinder, the cylinder's boundary oriented to TX is discretized to 135 candidate points shown in Fig. 4. (Spacing between two candidate points is $1°$, equivalent to a 0.8727-mm length on the arc.) The number of candidate points is fine enough to disclose the tendency of time variation between the source and the focal point. Fig. 5 shows the time elapsed between the source and the focal point via 135 possible refraction points, which is calculated by

$$T_N = \frac{d_{1T\_N}}{c} + \frac{\sqrt{\varepsilon_r} \cdot d_{2T\_N}}{c}, \quad N = 1 \sim 135 \quad (4)$$

where $c$ is the speed of light in air. The results indicate three local minima and maxima, each corresponding to a real path the wave would take as shown in Fig. 4. From the mathematic point of view, all of them are the solutions to the equation set as follows:

$$\begin{cases} x^2 + y^2 = r^2 (x > 0) \\ \dfrac{\sin[\theta_i(x, y)]}{\sin[\theta_r(x, y)]} = \sqrt{\varepsilon_r}. \end{cases} \quad (5)$$

The first equation is the circle equation for radii $r = 5$ cm. The second equation represents Snell's equation, where $\theta_i$ and $\theta_r$ are the incident angle in air and the refraction angle in the cylinder, respectively. $(x, y)$ is the coordinate of the refraction point on the border to be solved. For the specific focal point shown in Fig. 4, there are three valid solutions to (5), and the number of solutions may vary when the focal point switches. Fig. 6 shows the distribution of the number of solutions to (5) for all focal points in ROI. Black, dark gray, light gray, and white denote three, two, one, and zero paths for the wave to propagate from TX to focal point inside ROI, respectively. It is interesting to see that the first candidate point ($N = 1$) shown in Fig. 5, which represents an overall least time, is not a solution to (5), because it is not a "stationary" time length. In addition, despite no solutions to (5) for the white area, a "least time" can still be found among the 135 candidate points by solving (4), which is not a "stationary" solution either, and would be a source of clutters in the image if it was utilized in methods based on TS evaluations. Fig. 6 shows the assumption that microwaves behave like a ray. In the UWB spectrum (3.1–10.6 GHz), our electromagnetic simulation shows that the larger the dielectric constant of medium 2, and/or the higher the frequency, the more approximate (to the grayed area in Fig. 6) the field pattern in the cylinder is achieved. It was observed strong-magnitude fields appeared in the left–middle region of the cylinder, indicating that waves were refocusing in this place after they passed across the interface.

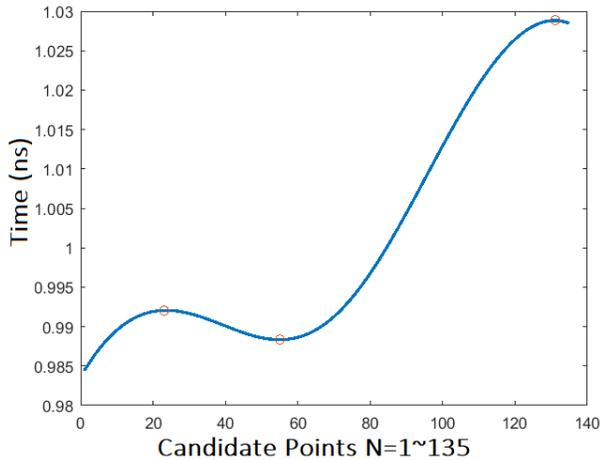

Fig. 5. Time elapsed as the wave propagates from the source to 135 candidate points on the cylinder's boundary and then to the focal point inside the cylinder shown in Fig. 4. There are three real refraction points which results in three propagation paths.

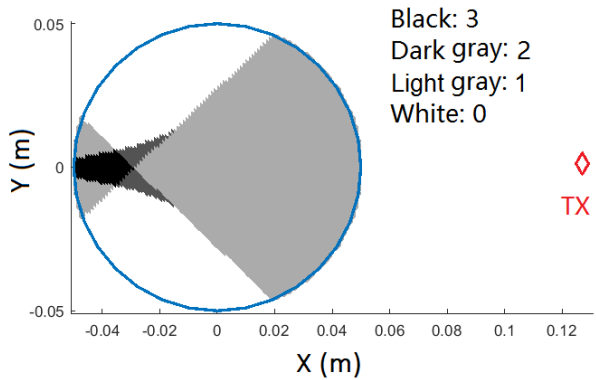

Fig. 6. Number of wave propagation paths at 4.5 GHz when the wave propagates from TX at (0.13 cm, 0) in the air into the cylinder with the permittivity of 6.5. There are three, two, one, and zero propagation paths from TX to a focal point inside black, dark gray, light gray, and white regions inside the cylinder, respectively. The distribution of these shaded regions is slightly different for other frequencies within the range 2–7 GHz.

For convenience, we assumed that the ray method was still valid to predict the propagation path for low-frequency components in the UWB spectrum. Referring to Fig. 6, there are up to three paths for a harmonic wave to propagate from TX to a certain focal point. Similarly, there can be up to three paths for a harmonic wave to propagate from the focal point to a receiver outside the cylinder. Hence, there will be, at most, nine propagation paths from TX to the focal point and then to the receiver. A signal collected by the receiver shall be the sum of the wave coming from all paths. More challenging, the phase of the received signal is determined by both the amplitude and phase of the signal from all path. Therefore, taking $|\vec{R}|$ and $G(x, y, f)$ into account of compensation is necessary, which differs from the traditional SAR.

The good thing is, unlike the multipath effect in a communication system, once the focal point and the antenna location are selected (assuming that the antenna's pattern mode is premeasured so $G_t(x, y, f)$ and $G_r(x, y, f)$ are known), all paths are predictable and each path can be assumed as a channel denoted by a transfer function $H$, which can be represented by a complex number whose magnitude indicates the decay and phase represents the PS of the channel. A signal sent by TX, passing through the same focal point via different channels and finally arriving to a receiver is like the signal passing a combined transfer function

$$H_{\text{eff}} = \sum_{i=1}^{n} H_i \qquad (6)$$

where $i$ and $n$ represent the index and the total number of the channels, respectively, and $H_{\text{eff}}$ is the effective combined transfer function. Therefore, for all the focal points having monochromatic multipaths to the antenna, the compensation term $(|\vec{R}|e^{j\vec{k}\vec{R}}/(G_t(x, y) \cdot G_r(x, y))^{1/2})$ in (2) shall be replaced by $(1/H_{\text{eff}})$. Equation (2) is used for those focal points having single path to TX while to the receiver. Although the ray-based channel-prediction method is derived from the far-field theory, we find it works well when exploring the problem as shown in Fig. 6, which will be demonstrated in Section IV.

## III. EXPERIMENTAL SETUP

The schematic of the UWB microwave imaging system we utilized is presented in Fig. 7(a), and the experimental setup is presented in Fig. 7(b). This system applies a pair of self-designed/fabricated unidirectional UWB antennas mounted on an inner circular ring with radius of 98 mm, and an outer circular ring with radius of 130 mm, to form a multistatic mode. Each antenna can rotate independently about the central table. Movements are accurately controlled by two-step motors and our self-developed software and hardware circuits. Fig. 8 shows the fabricated unidirectional UWB antenna and its $S_{11}$ coefficient, where the −6-dB return-loss bandwidth is from 1.5 to 7.7 GHz. The main body of the measurement system is made of polyvinyl chloride and wood to produce the low environmental reflection. More information about this antenna can be found in [19] and the entire measurement system in [20] and [21]. The object under test was hung in a plastic graduated cylinder (PGC) filled with glycerin. The vertical position of the object is approximately the same as the center of the two antennas. PGC with a diameter of 10 cm was ideally positioned in the center of the supporting table. The wall thickness of PGC is about 1 mm, which is ignored in our experiment. Glycerin is chosen as the background medium because it often serves as the coupling medium in biomedical microwave imaging experiments [22]–[25]. The solid lines shown in Fig. 9 reveal the permittivity and conductivity, respectively, of glycerin from 2 to 7 GHz at room temperature as measured by an Agilent dielectric probe kit AT-85070E. Its dispersive property, although seems weak, is not trivial when a typical UWB pulse for medical detection, such as the one shown in Fig. 10 is used, whose full-width at half-maximum (FWHM) is 0.146 ns and 3-dB bandwidth from 2 to 7 GHz. The TS difference between 2 and 7 GHz for a propagation distance $d$ can be approximated by

$$\Delta t = \frac{(\sqrt{\varepsilon_{7\,\text{GHz}}} - \sqrt{\varepsilon_{2\,\text{GHz}}})d}{c}. \quad (7)$$

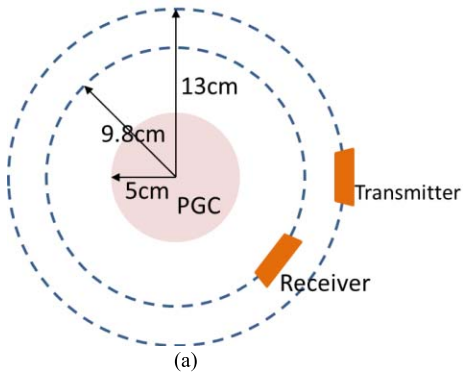

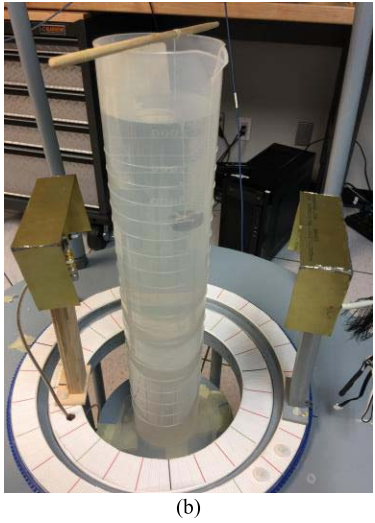

Fig. 7. (a) System schematic. (b) Photograph of the microwave imaging device and a metal object under test hung in PGC filled with glycerin.

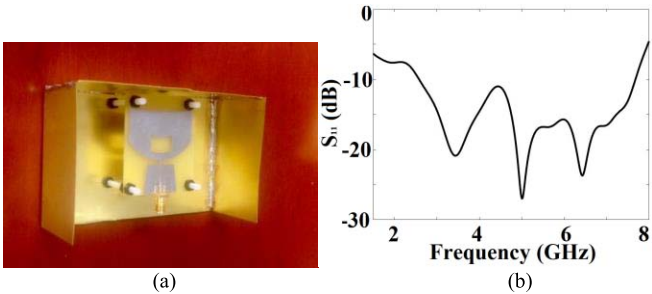

Fig. 8. (a) Fabricated unidirectional UWB antenna. (b) Measured $S_{11}$ of the antenna.

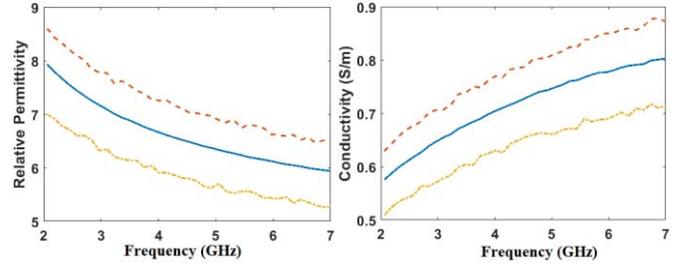

Fig. 9. Solid line shows the measured permittivity (left) and conductivity (right) of glycerin. -- represents 8%–12% randomly higher than the solid data, and ... represents 8%–12% randomly lower than the solid data.

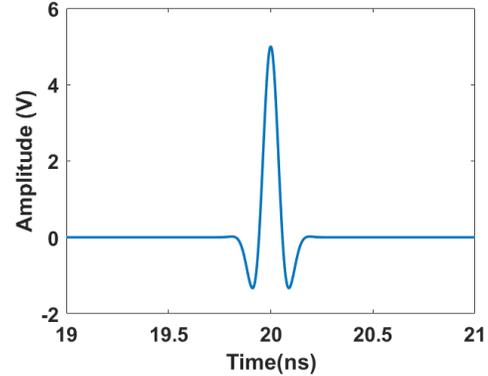

Fig. 10. Ideal pulse delivered to the port of TX in the experiment.

When $d = 10$ cm, (7) yields a 0.143-ns difference between 2 and 7 GHz when the UWB signal propagates in glycerin having a dielectric constant shown in Fig. 9, which is about 98% of FWHM of the UWB pulse.

Our measurement was performed in the time domain, and then the acquired data were converted to the frequency domain for further processing. During the measurement, TX and receiver were controlled to rotate about the cylinder in 15° increments and kept $\geq 45°$ apart to prevent potential couplings. For example, when TX is at 0°, the receiver starts to collect data from 45° and ends at 315°. Then, TX moves to 15°, and the receiver will return to 60° to start the collection, until ending at 330°. With this setup, $n$ in (6) can only be 2, 3, or 4 ($n = 6$ or 9 will not exist) so the complexity of computation is reduced. We used Tektronix arbitrary wave generator (AWG) 70001A (sample rate 50 GSa/s) to produce the source pulse, which is then amplified by a mini-circuit ZVA-183W + amplifier (effective gain 25 dB from 100 MHz to 18 GHz) before it is fed to TX antenna. The signal fed to the antenna is the one shown in Fig. 10 (loss and distortion in the cable and amplifier have been calibrated). The signal acquired by the receiver was sampled and digitized by a Tektronix digital phosphor oscilloscope (DPO) 71604C with a sample rate of 100 GSa/s. DPO was set to make six measurements (the number of measurements is adjustable in our software) and save the average for every acquisition position to reduce random errors. Once data are successfully saved on DPO, the software informs the antenna to move to the next position. The entire experiment, including antenna rotations, AWG, and oscilloscope triggering, is fully automated as long as all required parameters in the software are correctly set at the beginning of the experiment.

Alternatively, a measurement can be conducted in the frequency domain using a VNA [16]. Our measurement system and software support both time-domain and frequency-domain scans. In order to compare PSAS with the conventional TS-based algorithms, we chose to perform the measurement in the time domain.

## IV. Image Reconstruction

### A. Artifact Removal

Ideally, the scattered field $E_{\text{scat}}$ of the object under test can be obtained by

$$E_{\text{scat}} = E_{\text{total}} - E_{\text{pre}} \tag{8}$$

where $E_{\text{pre}}$ is a premeasured field when there is no object present in PGC and $E_{\text{total}}$ is the total field when the object presents in PGC. The subtraction can provide the perfect object scattering signal. However, $E_{\text{pre}}$ is sometimes difficult to be accessed in practice. The average subtraction method is more realistic, in which the artifact is estimated as an average of the signal recorded in each channel and is, then, removed by subtracting from each received signal

$$V_{\text{scat}}^i = V_{\text{total}}^i - \frac{1}{N}\sum_{i=1}^{N} V_{\text{total}}^i. \tag{9}$$

In our specific problem, the artifact is composed of incident signals, air–PGC interface reflection, and PGC–glycerin interface reflection. As performing this method to our multistatic mechanism, according to the symmetry of the system, it can be supposed that the artifact received by receiver at 45° and TX at 0°, recorded as $T_0 R_{45}$, shall be equivalent to the artifact when the receiver is at 60° and TX at 15°, recorded as $T_{15} R_{60}$, i.e., TX and receiver having the same relative positions. Therefore, we classified the signals into 19 groups, and each group contains 24 data sets

$$19\text{ groups}\begin{cases} T_0 R_{45},\ T_{15} R_{60},\ \cdots\ T_{345} R_{30}; \\ T_0 R_{60},\ T_{15} R_{75},\ \cdots\ T_{345} R_{45}; \\ \quad\quad\quad\quad \vdots \\ T_0 R_{315},\ T_{15} R_{330},\ \cdots\ T_{345} R_{300}. \end{cases}$$

$$\underbrace{\hspace{4cm}}_{24\text{ datasets}}$$

In each group, the object scattering signal is obtained by using (9) when set $N = 24$. Therefore, a complete measurement only requires collecting the total field, which is finished within 10 min.

### B. Reconstruction and Comparison

Even if the PSAS algorithm may solve the dispersion problem, PS may still not be precisely evaluated caused by two main reasons: the inhomogeneity of the background medium and deficient dielectric knowledge of the background medium (inconsistence with the reference data). We consider using the two dielectric data sets shown in Fig. 9(a) and (b) as the reference to evaluate the PS and amplitude decay in the PSAS reconstruction: the first one is 8%–12% randomly higher than the measured data (at each frequency) and the second is 8%–12% randomly lower than the measured data. This assumption is rational because the dielectric parameters of the medium are more often seen exclusively higher or lower than the reference (for example, due to water content). The case that the line of the real parameter and the reference data present a cross is rare. So, by using the two data sets, we investigate the performance of PSAS when PS and amplitude decay are not accurately estimated.

After the time-domain signals were converted to the frequency domain, we picked up 11 frequency points from 2 to 7 GHz with a 0.5-GHz increment to run PSAS for the reason of efficiency. The phase center of the fabricated antenna and the PS from the phase center to the antenna's port were premeasured at these frequencies. The first object under test was composed of eight stacked one-cent U.S. coins tied up together, as shown in Fig. 11(s). Fig. 11(a)–(i) shows the 2-D reconstructed image by PSAS, RAR, and DMAS when the metal object was intentionally placed close to the wall of PGC, which place belongs to the monochromatic-multipath area for certain antennas. The second object under test was a rectangular plasticine with dimension $17 \times 10 \times 12$ mm as shown in Fig. 11(v), whose main components are clay and aliphatic acids. Plasticine is a weak scatterer at microwave frequencies. The reconstructions for the plasticine are shown in Fig. 11(j)–(r). Each image took about 2 min to achieve on an i-7 Intel CPU. We selected DMAS for comparison because it is a classic and very effective approach in microwave near-field imaging. The reason RAR was chosen is that it provides high contrast and high SNR, which is accomplished by adding a weight to the sum of signals for each focal point. Since in the object's location, the weight (computed by the correlation of neighbor antennas) applied is much larger than elsewhere not presenting an object, pixel strength in the object's location can be further enhanced. A detailed elaboration of how to calculate the added weight when a monostatic collection is utilized is provided in [15]. As far as the experimental setup depicted in Section III is concerned, we need to extend this method for a multistatic collection mode: First, we used the same method as in [15] to compute the signal correlation between neighbor receivers within one transmission. Next, the correlation was used to compute the weight which is, then, multiplied with the sum of the signal within the same transmission. These steps were repeated in every transmission. Mathematically,

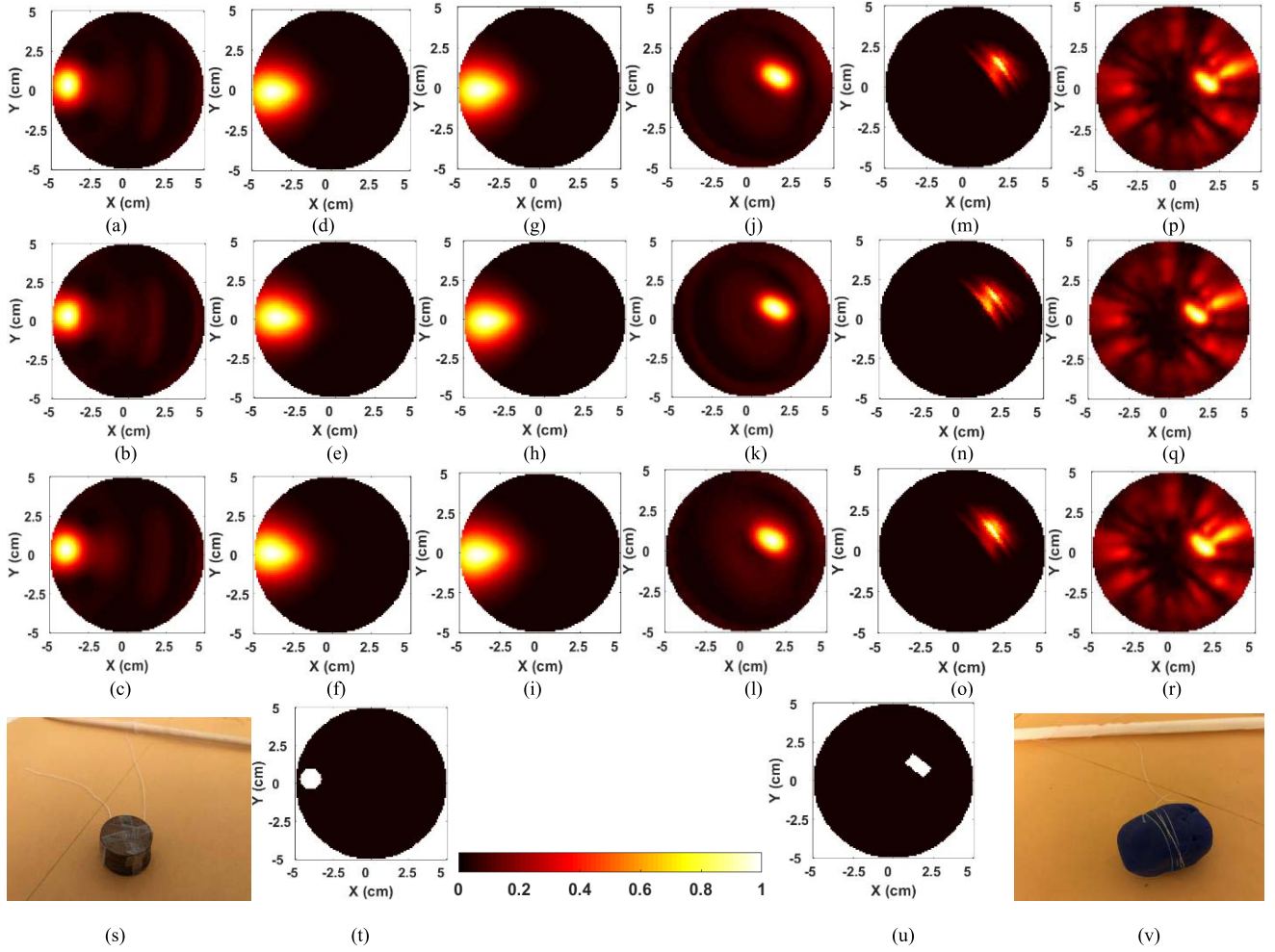

Fig. 11. Bright area in (a)–(i) represents the metal object in the reconstructed image. The dielectric parameter used in PSAS is (a) equal to, (b) 8%–12% higher than, and (c) 8%–12% lower than measured data. The dielectric parameter used in RAR is (d) equal to, (e) 8%–12% higher than, and (f) 8%–12% lower than measured data at 4.5 GHz. The dielectric parameter used in DMAS is (g) equal to, (h) 8%–12% higher than, and (i) 8%–12% lower than measured data at 4.5 GHz. The bright area in (j)–(t) represents the plasticine object in the reconstructed image. The dielectric parameter used in PSAS is (j) equal to, (k) 8%–12% higher than, and (l) 8%–12% lower than measured data. The dielectric parameter used in RAR is (m) equal to, (n) 8%–12% higher than, and (o) 8%–12% lower than measured data at 4.5 GHz. The dielectric parameter used in DMAS is (p) equal to, (q) 8%–12% higher than, and (r) 8%–12% lower than measured data at 4.5 GHz. (s) and (v) Object under test. (t) and (u) Ideal reconstructed image for two objects, respectively.

the modified RAR approach for a multistatic measurement is

$$I = \int \sum_{i=1}^{M} \left( w_i \cdot \sum_{j=1}^{N} V_{ij}(t + \Delta t) \right)^2 \cdot dt \quad (10)$$

where $M$ is the number of TXs, $N$ is the number of receivers, and $w_i$ represents the weight in the $i$th transmission. $I$ will be the pixel value in the image.

Signal propagation speed used to compute TS in RAR and DMAS is under the assumption that glycerin has a constant permittivity over the entire UWB spectrum at the center frequency (4.5 GHz). The least-time method was still applied to the path estimation in RAR and DMAS because time-domain signals are unable to be processed by (6) even though the monochromatic multipaths are solved for 4.5 GHz. It is observed that all the three methods can localize both the strong scatterer and the weak scatterer correctly in different positions in ROI and all are robust to an 8%–12% background dielectric inaccuracy. However, since the dispersive and monochromatic multipath issues were not considered in RAR and DMAS, the profile of the reconstructed object is distorted from its original shape although the image contrast made these two methods may be even higher than PSAS. To quantitatively compare the reconstructed images, SNR defined by

$$\text{SNR} = 20 \cdot \log_{10} \frac{\text{Max}(I_r)}{\text{Ave}(I_r)} \quad (11)$$

was utilized to represent the ratio of the strongest to the average pixel value in the image, where $I_r$ is the pixel value, Max represents a maximum, and Ave denotes an average. In addition, the image contrast, calculated by the ratio of the average pixel value in the object's region $\Omega$ to the average value of the entire image

$$\text{Ctr} = 20 \cdot \log_{10} \frac{\frac{1}{N} \sum_{\Omega}(I_r)}{\text{Ave}(I_r)} \quad (12)$$

was adopted, where $N$ is the number of pixels in the object's region $\Omega$ which is defined by $I_r \geq (1/2)\text{Max}(I_r)$. Both SNR and contrast for the reconstructed images shown in Fig. 11 are summarized in Table I. For the metal object, PSAS provides better SNR and contrast than RAR and DMAS. For the weak scatterer, RAR provides the highest SNR and contrast, but RAR seems over-enhanced some pixels within the object's region, leading to the unclearness of the object's edge. Therefore, SNR and image contrast might not be sufficient to make a convincible quantitative comparison. Table I also presents the relative difference $\delta$ between the reconstructed images and an ideal object profile, which is calculated by

$$\delta = \frac{\iint_D [I_r(x,y) - I(x,y)]^2 \cdot ds}{\iint_D [I(x,y)]^2 \cdot ds} \quad (13)$$

where $(x, y)$ is the coordinate of the pixel in ROI $D$ and $I(x, y)$ represents the pixel value at $(x, y)$ in the ideal profile image. It is found that the smallest relative difference was obtained by the PSAS method in all cases. This result indicates that the PSAS image effectively reduces the distortion with respect to the object. SNR, Ctr, and $\delta$ (when accurate dielectric parameters are applied to reconstruction) are also presented visually in Fig. 12 for better understanding the quantitative results shown in Table I.

TABLE I
QUANTITATIVE COMPARISON OF THREE RECONSTRUCTION METHODS

| Method | and Figure | SNR (dB) | Ctr (dB) | δ |
|---|---|---|---|---|
| PSAS | Fig. 11 (a) | 23.6933 | 20.7200 | 0.4636 |
| PSAS | Fig. 11 (b) | 23.4035 | 20.4571 | 0.4835 |
| PSAS | Fig. 11 (c) | 23.6085 | 20.6744 | 0.4809 |
| RAR | Fig. 11 (d) | 21.8854 | 18.8584 | 1.2357 |
| RAR | Fig. 11 (e) | 21.6645 | 18.7269 | 1.2458 |
| RAR | Fig. 11 (f) | 21.4078 | 18.5533 | 1.2628 |
| DMAS | Fig. 11 (g) | 21.8497 | 18.8270 | 1.2640 |
| DMAS | Fig. 11 (h) | 21.6297 | 18.7192 | 1.3554 |
| DMAS | Fig. 11 (i) | 21.3839 | 18.5330 | 1.3842 |
| PSAS | Fig. 11 (j) | 23.3676 | 20.6082 | 0.7028 |
| PSAS | Fig. 11 (k) | 23.3546 | 20.5739 | 0.7548 |
| PSAS | Fig. 11 (l) | 23.3606 | 20.5817 | 0.8225 |
| RAR | Fig. 11 (m) | 31.1871 | 28.2152 | 1.0090 |
| RAR | Fig. 11 (n) | 30.8156 | 26.9024 | 1.0309 |
| RAR | Fig. 11 (o) | 30.8365 | 27.5449 | 1.1212 |
| DMAS | Fig. 11 (p) | 18.8163 | 15.2845 | 2.5294 |
| DMAS | Fig. 11 (q) | 18.5445 | 15.1174 | 2.7080 |
| DMAS | Fig. 11 (r) | 18.1969 | 14.8610 | 2.9264 |

In the third test, we placed the metal object and the plasticine object in GPC simultaneously and applied three methods to reconstruct image using the dielectric parameters shown by dotted line (8%–12% lower than the referenced standard) shown in Fig. 9. The reconstructions are shown in Fig. 13. The brightest region in three images indicates the metal object. By further enhancing the value of the strongest pixels to increase contrast, RAR totally missed the weak scatterer, which was expected to present in the white frame. DMAS is a little bit better than RAR in finding the weak scatterer but is not as good as PSAS. Therefore, although RAR and DMAS do suppress the background noise well, they might miss weaker scatters when multiple objects present.

To summarize the advantages of PSAS and compare them with RAR and DMAS, Table II presents the comparison from three points of view: image quality based on SNR, Ctr, and $\delta$; probability of missing weak scatters when multiple scatters exist; and the extent of distorting the object shape.

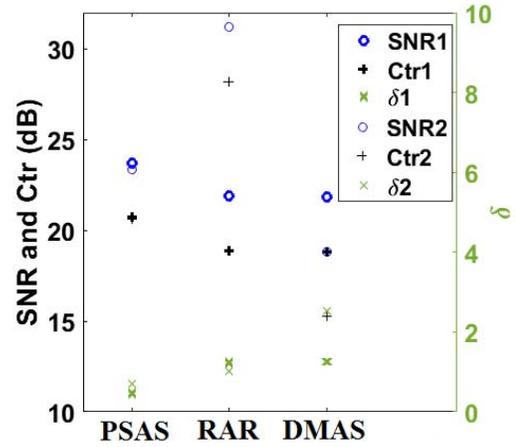

Fig. 12. Comparison of the data shown in Table I for the case using accurate dielectric parameters. SNR1, Ctr1, and $\delta 1$ are SNR, contrast, and relative difference for the coin's reconstructed image, and SNR2, Ctr2, and $\delta 2$ are SNR, contrast, and relative difference for the plasticine object's reconstructed image.

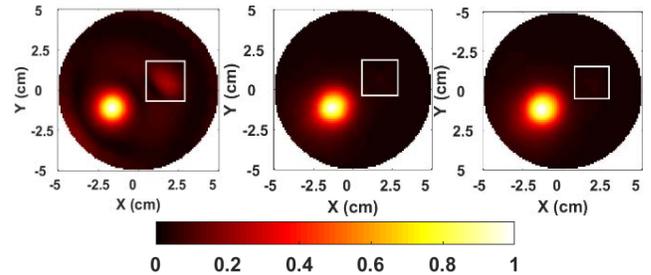

Fig. 13. Strong scatter and weak scatter concurrently present in PGC. Reconstruction was made by PSAS (left), RAR (middle), and DMAS (right), respectively.

TABLE II
QUALITATIVE COMPARISON OF THREE RECONSTRUCTION METHODS

| Method | Image Quality | Missing Weak Scatterer | Distorting Object |
|---|---|---|---|
| PSAS | Good | Lowest Possibility | Lowest Possibility |
| RAR | Good | Probably | High Possibility |
| DMAS | Poor | Probably | Low Possibility |

Keep in mind that only 11 frequencies were adopted in composing the PSAS images. If more frequencies are applied, better images can be expected by PSAS at cost of longer computational time. According to the reconstruction shown in Fig. 11(a)–(i), one might notice that PSAS potentially provides better resolution. This can be explored by observing the target response signal used by the three methods. The dashed line shown in Fig. 14 illustrates the TS metal target response (single-object case shown in Fig. 11) when TX is at 0° and the receiver is at 180°, which is used by RAR and DMAS to form images. The solid-line signal, obtained by first converting the target response into the frequency domain, picking 901 frequency points in the range from 1 to 10 GHz with a 10-MHz interval, then phase-shifted and compensated at these frequencies and finally converting back to the time domain, represents the PS target response.

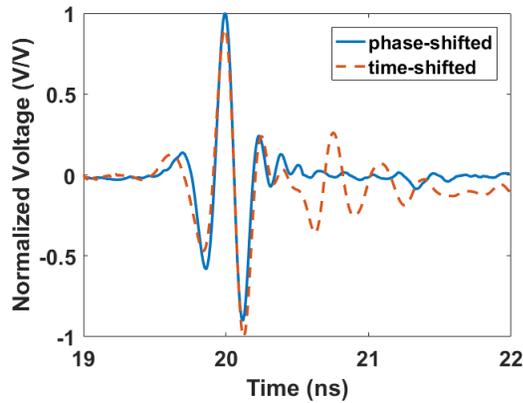

Fig. 14. Comparison of the target response signal after a TS process and a PS process in the time domain.

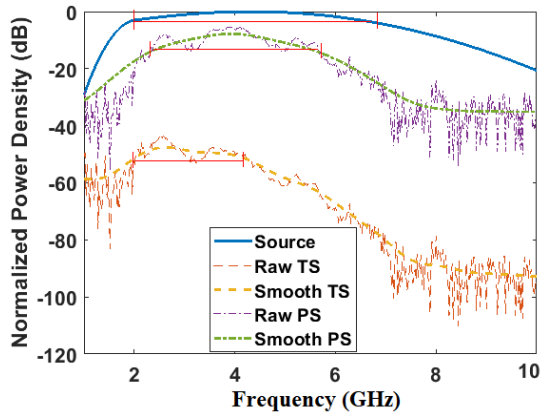

Fig. 15. Spectrum comparison of the source signal with the target response after a TS process and a PS process. The source signal is the measured data at the port of TX.

Two signals are normalized to make an easy comparison. The TS signal has a wider main pulse, resulting from the medium dispersion, which can be expected. The TS signal also has more ripples, probably resulting from the multipath effect, since the object is located or partly located in the black area shown in Fig. 6. Therefore, from the point of view in the time domain, PSAS may optimize the waveform of the UWB target response signal, thus improve the image. Note that the PSAS reconstruction does not need to convert the frequency response back to the time domain. This procedure is conducted in the present analysis for obtaining a nice time-domain signal to be compared with the TS signal. Also, in the real image reconstruction, fewer frequencies are needed to acquire an image; e.g., the PSAS images shown in Figs. 11 and 12 are obtained by using 11 frequencies only.

Comparison may also be conducted in the frequency domain. Fig. 15 presents the spectrum of the normalized source signal, object response after a TS compensation, and object response after a PS compensation. The source signal is the measured data at the port of TX. Because it is uneasy to extract the feature from the raw responses, a Gaussian-weighted moving average window was applied to smooth the data. Then, a 3-dB bandwidth of each smoothed signal was marked. Fig. 15 indicates that since the high-frequency components decay more rapidly, the bandwidth (TS signal) is significantly reduced compared to the original source. Furthermore, the center frequency has deviated to a lower place from 4.5 GHz (source center frequency). Thus, the validity of still using velocity at 4.5 GHz in TS algorithms becomes doubtful. Moreover, in a complete measurement, signal received by different antennas passing through different length of paths causes different decays. So, each received signal has a different center frequency. This makes more challenging to determine the velocity to be used in the time delay evaluation. With the PS compensation, it can be seen in Fig. 15 that the high-frequency components were given more compensation, so the bandwidth is effectively recovered, which will contribute to the resolution improvement.

## V. CONCLUSION

A new algorithm for UWB radar imaging in dispersive media, called PSAS, is presented. This algorithm compensates the PS and amplitude attenuation for each frequency component in the UWB signal individually and finally sums the contribution from all frequency components. This algorithm has better potential to overcome the multispeed and multipath issue when UWB signals propagate in dispersive media than traditional TS methods. In addition, the algorithm takes the monochromatic multipath effect due to refraction on a curved boundary into account. The experimental results and comparisons between PSAS and two other algorithms validate the algorithm and the prediction of the best performance. The optimization of the pulse waveform as the UWB signal propagates in a dispersive medium indicates the superiority of PSAS over TS methods in object detection and image reconstruction.